\documentclass[epj,twocolumn]{webofc}

\usepackage{bm}
\usepackage[varg]{txfonts}   
\usepackage[export]{adjustbox}

\newcommand{\vv}{\bm{v}}
\newcommand{\cc}{\bm{c}}
\newcommand{\s}{\widehat{\bm{\sigma}}}
\newcommand{\hcs}{\mathrm{H}}
\newcommand{\mx}{\mathrm{M}}
\newcommand{\rref}{\mathrm{R}}
\newcommand{\dif}{d}

\begin{document}
\title{Relative entropy of freely cooling granular gases. A molecular dynamics study}
%
%

\author{\firstname{Alberto} \lastname{Meg\'ias}\inst{1}\fnsep\thanks{\email{albertom@unex.es}} \and
        \firstname{Andr\'es} \lastname{Santos}\inst{1,2}\fnsep\thanks{\email{andres@unex.es}}
}

\institute{Departamento de F\'isica, Universidad de Extremadura, E-06006 Badajoz, Spain
\and
Instituto de Computaci\'on Cient\'ifica Avanzada (ICCAEx), Universidad de Extremadura, E-06006 Badajoz, Spain
          }

\abstract{
    Whereas the original Boltzmann's $H$-theorem applies to elastic collisions, its rigorous generalization to the inelastic case is still lacking. Nonetheless, it has been conjectured in the literature that the relative entropy of the velocity distribution function with respect to the homogeneous cooling state (HCS) represents an adequate  nonequilibrium entropy-like functional for an isolated freely cooling granular gas. In this work, we present molecular dynamics results reinforcing this conjecture and rejecting the choice of the Maxwellian over the HCS as a reference distribution. These results are qualitatively predicted by a simplified theoretical toy model. Additionally, a Maxwell-demon-like velocity-inversion simulation experiment highlights the microscopic irreversibility of the granular gas dynamics, monitored by the relative entropy, where a short ``anti-kinetic'' transient regime appears for nearly elastic collisions only.
}
\maketitle

\section{Introduction}
\label{intro}
Granular gases are modeled in their simplest form as inelastic hard spheres with a constant coefficient of restitution, $\alpha$ \cite{D00,
G03,
BP04,
G19}. It is well known that granular gases are intrinsically  out of equilibrium and that a description by means of kinetic theory is meaningful. In a kinetic-theoretical description of a granular gas, one defines the \emph{granular} temperature as the mean kinetic energy per particle, as an analogue to its definition for molecular gases. Even though this temperature is not a \emph{thermodynamic} temperature, one can look for the nonequilibrium entropy-like functional of this system, i.e., a Lyapunov functional, in analogy with Boltzmann's $H$-functional and the celebrated $H$-theorem for elastic collisions

The problem introduced and solved by Boltzmann in 1872 \cite{B95} is not easy to extend in the context of granular gases and the associated inelastic form of the renowned Boltzmann equation. The original $H$-functional is precisely the Shannon measure \cite{S48} of the one-particle velocity distribution function \cite{CC70,GS03}. However, it is known that, in its continuous description, Shannon's entropy presents the so-called \emph{measure} problem \cite{MT11}, i.e., it does not weigh properly the phase space. In the elastic case, this problem is easily solved by considering the relative entropy (or Kullback--Leibler divergence \cite{KL51}) of the one-particle velocity distribution function with respect to the Maxwellian distribution, which becomes the original $H$-functional up to a constant in that case. Moreover, some relevant properties of the elastic-particle system, like collisional symmetry and reversibility, do not hold anymore in the inelastic scheme. Then, the proper entropy-like functional must solve these issues.

The quest of such a quantity in the homogeneous case has been addressed mathematically in Refs.\ \cite{MMR06,MM06,MM09} in the context of the inelastic Boltzmann equation, and in Ref.\ \cite{GMMMRT15} from a stochastic point of view. Both approaches converge into a single functional, which is proved in the \emph{quasielastic} limit, i.e., $1-\alpha \ll 1$, to be the entropy-like functional associated with this system. In the case of free cooling, the conjectured quantity is the relative entropy of the reduced velocity distribution function, $\phi$, with respect to the homogeneous cooling state (HCS), $\phi_{\hcs}$, chosen as the proper reference distribution. This conjecture was recently reinforced with computer simulations in the whole range of inelasticity \cite{MS20}.

In this work, we complement the study carried out in Ref.\ \cite{MS20} with new simulations. First, we study the problem by means of a simplified \emph{toy model} \cite{MS20} and  investigate how it highlights the possible Lyapunov character of the proposed functional for two different reference distributions, namely the Maxwellian and the HCS distributions. Next, molecular dynamics (MD) results are presented and compared with the predicted theoretical behavior, including three systems not considered in Ref.\ \cite{MS20}. Finally, as a fully original contribution of this work, we  report MD results for a sort of Maxwell-demon experiment where the irreversibility of the collisional process and the possibility of an anti-kinetic stage are discussed.

\section{A toy model}
\label{sec-1}
Let us consider a granular-gas  model of inelastic and smooth hard spheres with collisional rules

\begin{equation}\label{CR}
    \vv_{1,2}^{\prime\prime} = \vv_{1,2}\mp\frac{1+\alpha}{2\alpha}(\vv_{12}\cdot\s)\s,\quad \vv_{12}''\cdot\s=-\frac{1}{\alpha}\vv_{12}\cdot\s,
\end{equation}
for the precollisional velocities, where $\vv_{12}$ is the relative velocity, and $\s$ the unit intercenter vector. We will assume that the system satisfies the inelastic Boltzmann equation, which in reduced units reads
\begin{equation}\label{BE}
    \frac{\kappa}{2}\partial_s\phi(\cc;s)+\frac{\mu_2(s)}{3}\frac{\partial}{\partial \cc}\cdot\left[\cc \phi(\cc;s)\right]=I[\cc|\phi,\phi].
\end{equation}
Here, $\kappa \equiv 2\sqrt{2\pi}$ is a constant, $\cc=\vv/v_{\mathrm{th}}$ is the reduced velocity, $v_{\mathrm{th}}(t)=\sqrt{2T(t)/m}$ is the thermal velocity, $m$ is the mass of a particle, $T=m\langle v^2\rangle /3$ is the granular temperature, which decreases monotonically following Haff's law \cite{H83,BP04,G19,BE98}, $s=\frac{1}{2}\int_0^t \medspace {d}t^\prime \nu(t^\prime)$ is the (nominal) average number of collisions per particle up to time $t$, where $\nu(t)=\kappa n \sigma^{2}v_{\mathrm{th}}(t)$  is the collision frequency, $I[\cc|\phi,\phi]$ is the collisional operator in reduced units, and $\mu_2=-\int\dif\cc\medspace c^2 I[\cc|\phi,\phi]$ is  the reduced cooling rate.

The relative entropy, or Kullback--Leibler divergence, of a velocity distribution, $\phi$, with respect to a reference distribution, $\phi_{\rref}$, is defined as
\begin{equation}\label{DKL1}
    \mathcal{D}_{\mathrm{KL}}\left(\phi\|\phi_{\rref}\right) = \int{d}\cc\medspace \phi(\cc)\ln \frac{\phi(\cc)}{\phi_{\rref}(\cc)}.
\end{equation}
This functional  is convex, non-negative, and identically zero if and only if $\phi = \phi_{\rref}$ \cite{KL51}.

We assume that both $\phi$ and $\phi_{\rref}$ are isotropic and can be expanded around the Maxwellian $\phi_{\mx}(\cc)=\pi^{-3/2}e^{-c^2}$ in terms of Sonine polynomials,
\begin{equation}\label{SonineSum}
    \phi(\cc;s) = \phi_{\mx}(\cc)\left[1+\sum_{k= 2}^\infty a_k(s) S_k(c^2)\right],
\end{equation}
where $S_k$ is the $k$-th Sonine polynomial and $a_k$ is the $2k$-th cumulant of the distribution, defined as $a_k=\langle S_k\rangle/\mathcal{N}_k$ with  $\mathcal{N}_k=(2k+1)!!/2^k k!$. By definition, $a_0=1$ and $a_1=0$, so that the first nontrivial coefficient is the fourth cumulant $a_2=\frac{4}{15}\langle c^2\rangle-1$.

Let us now construct a  toy model of $\mathcal{D}_{\mathrm{KL}}\left(\phi\|\phi_{\rref}\right)$   \cite{MS20} for an arbitrary reference distribution. Imagine a \emph{perturbative} parameter $\varepsilon$ in front of the Sonine summation in Eq.\  \eqref{SonineSum}. Expanding in powers of $\varepsilon$ and keeping terms up to second order, we have
\begin{align}\label{DKL2}
\frac{\phi(\cc;s)}{\phi_{\mx}(\cc)}\ln \frac{\phi(\cc;s)}{\phi_{\rref}(\cc)} =&
    \varepsilon \sum_{k=2}^\infty\Delta_{\rref}a_k(s)S_k(c^2)\nonumber \\
    &+\frac{\varepsilon^2}{2}\sum_{k= 2}^\infty\sum_{k^\prime= 2}^\infty\Delta_{\rref}a_{k}(s)\Delta_{\rref}a_{k'}(s)
   \nonumber\\
    &\times  S_k(c^2)S_{k^\prime}(c^2)+\mathcal{O}(\varepsilon^3),
\end{align}
where $\Delta_{\rref} a_k(s)\equiv a_k(s)-a_k^{\rref}$, $a_k^{\rref}$ being the Sonine coefficients for the reference distribution function $\phi_{\rref}(\cc)$. Inserting this expression into Eq.\ \eqref{DKL1} and using the orthogonality condition of the Sonine polynomials,  one obtains
\begin{subequations}
\begin{align}
    \mathcal{D}_{\mathrm{KL}}(\phi\|\phi_{\rref}) &= \frac{\epsilon^2}{2}\sum_{k= 2}^\infty \mathcal{N}_k [\Delta_{\rref} a_k(s)]^2+\mathcal{O}(\varepsilon^3), \label{DKLtoy} \\
    \partial_s \mathcal{D}_{\mathrm{KL}}(\phi\|\phi_{\rref}) &=\epsilon^2 \sum_{k= 2}^\infty \mathcal{N}_k \Delta_R a_k(s)\partial_s a_k(s)+\mathcal{O}(\varepsilon^3).\label{DKLtoy_der}
\end{align}
\end{subequations}
The r.h.s of Eq.\ \eqref{DKLtoy} is non-negative, and it is zero if and only if $a_k=a_k^{\rref}$ $\forall k \geq 2$, that is, $\phi=\phi_{\rref}$, in accordance with the properties of the relative entropy.
Next, we neglect terms of $\mathcal{O}(\varepsilon^3)$, formally take $\varepsilon=1$, and, as usually done in the literature \cite{BP04,G19,GS95,BRC96,vNE98,BP06,BP06b,MS00,SM09}, discard terms with $k\geq 3$. The result is
\begin{subequations}
\begin{align}
    \mathcal{D}_{\mathrm{KL}}(\phi\|\phi_{\rref})\approx& \frac{15}{16}[\Delta_{\rref}a_2(s)]^2, \label{DKLtoy_a2}
        \\
    \partial_s \mathcal{D}_{\mathrm{KL}}(\phi\|\phi_{\rref})\approx& -\frac{15}{8}K[1+a_2(s)]\Delta_{\rref}a_2(s)\Delta_{\hcs}a_2(s).\label{DKLtoy_a2_der}
\end{align}
\end{subequations}
Here, in consistency with the neglect of $a_k(s)$ for $k\geq 3$, we have used the evolution equation $\partial_s a_2(s)=-K [1+a_2(s)]\Delta_{\hcs}a_2(s)$ for the fourth cumulant \cite{MS20}, where  $\Delta_{\hcs} a_2(s)\equiv a_2(s)-a_2^{\hcs}$ and $K$ is a positive constant. Whereas the r.h.s of Eq.\ \eqref{DKLtoy_a2} is non-negative, the sign of the r.h.s of Eq.\ \eqref{DKLtoy_a2_der} is determined by the relative signs of $\Delta_{\rref}a_2(s)$ and $\Delta_{\hcs}a_2(s)$.

Let us consider two different reference distributions: the Maxwellian velocity distribution function, $\phi_{\mx}$, and the HCS velocity distribution function, $\phi_{\hcs}$. In the first case ($\rref=\mx$), one has $\Delta_{\mx} a_2(s)=a_2(s)$, and, therefore, $\partial_s \mathcal{D}_{\mathrm{KL}}(\phi\|\phi_{\mx})\leq 0$  only if either $a_2(s)\geq\max\{a_2^\hcs,0\}$ or $a_2(s)\leq\min\{a_2^\hcs,0\}$; conversely, $\partial_s \mathcal{D}_{\mathrm{KL}}(\phi\|\phi_{\mx})\geq 0$  only if either $0\leq a_2(s)\leq a_2^\hcs$ or $a_2^\hcs\leq a_2(s)\leq 0$.  Thus, our toy model shows that a monotonic relaxation of $\mathcal{D}_{\mathrm{KL}}(\phi\|\phi_{\mx})$ is not guaranteed. Let us assume, for instance,  that the initial value $a_2(0)$ is negative and $\alpha\lesssim 0.71$, so that the steady-value $a_2^{\hcs}$ is positive  \cite{BRC96,MS00,SM09,BP06,BP06b,MS20}; due to Bolzano's theorem,   during its evolution $a_2(s)$ must cross the zero value, so that  $\mathcal{D}_{\mathrm{KL}}(\phi\|\phi_{\mx})$ would present a local minimum. Analogously, a local minimum of $\mathcal{D}_{\mathrm{KL}}(\phi\|\phi_{\mx})$ is predicted by the toy model if $a_2(0)>0$ and $\alpha\gtrsim 0.71$, i.e., $a_2^{\hcs}<0$. In the case $\rref=\hcs$, however, $-\partial_s \mathcal{D}_{\mathrm{KL}}(\phi\|\phi_{\hcs})\propto [\Delta_\hcs a_2(s)]^2$ and the Lyapunov condition $\partial_s \mathcal{D}_{\mathrm{KL}}(\phi\|\phi_{\hcs})\leq 0$ is fulfilled.

\section{Molecular dynamics simulations}
\label{sec-2}

In order to check the predictions of the toy model for the two considered reference distributions, we have performed MD simulations using the DynamO software \cite{BSL11} for this model of granular gases. It is well known that the free cooling of granular gases presents long-wavelength instabilities \cite{G19,BDKS98}. In order to avoid them, we have simulated systems formed by $N=1.35\times 10^4$ particles in a cubic box of side length  $L/\sigma = 407.16$, which is at least $30$ times smaller than the critical length for the development of instabilities, which are not observed.

Initially, all particles  are arranged in an ordered crystalized configuration from which the system melts. The initial velocities are oriented along randomized directions with either a common magnitude (initial distribution $\delta$) or with a magnitude drawn from a Gamma ($\Gamma$) distribution. The respective initial values of the fourth cumulant are  $a_2(0)=-0.4$ ($\delta$ distribution) and $a_2(0)=0.4$ ($\Gamma$ distribution). Thus, according to the toy model, a nonmonotonic relaxation of $\mathcal{D}_{\mathrm{KL}}(\phi\|\phi_{\mx})$  is expected  for the initial distribution $\delta$ if $\alpha\lesssim 0.71$ and for the initial distribution $\Gamma$ if $\alpha\gtrsim 0.71$.

\begin{figure}[ht]
\centering
\includegraphics[width=8.cm,clip]{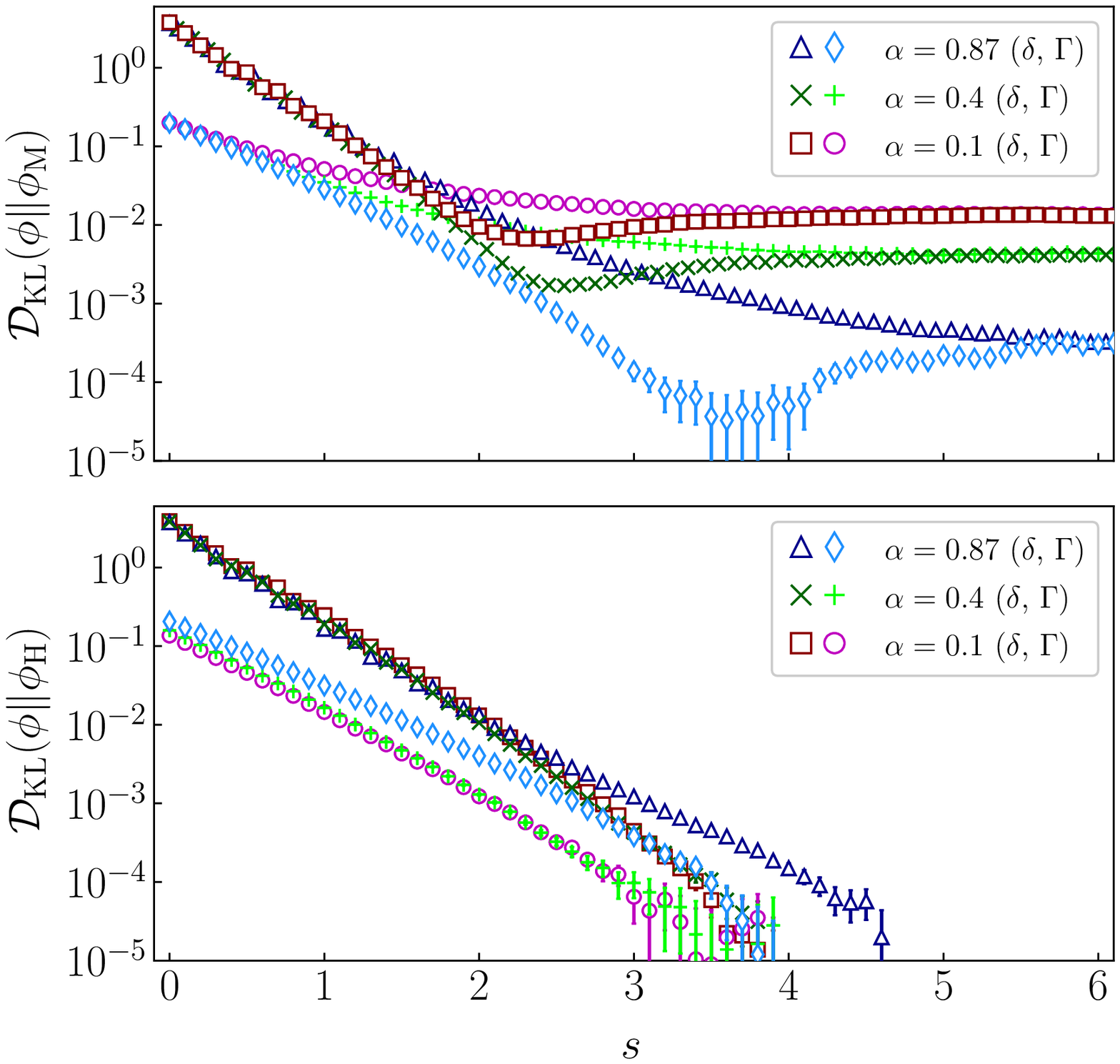}
\caption{
Evolution of  $\mathcal{D}_{\mathrm{KL}}(\phi\|\phi_{\rref})$ (in logarithmic scale) for $\rref=\mx$ (top panel) and $\rref=\hcs$  (bottom panel).
Symbols represent MD simulation results  for  coefficients of restitution  $\alpha=0.1$, $0.4$, and $0.87$, starting from the initial conditions
$\delta$ and $\Gamma$. The error bars are smaller than the size of the symbols, except  when
$\mathcal{D}_{\text{KL}}(\phi\|\phi_{\rref})\lesssim 10^{-4}$. Symbols for $s\gtrsim 5$ are discarded in the bottom panel because they appear as
purely noise, representing values that are out of the precision of the numerical scheme.}
\label{fig-1}       
\end{figure}

In Fig.\ \ref{fig-1} one can observe that, as predicted by the toy model, a local minimum  is actually observed during the evolution of $\mathcal{D}_{\mathrm{KL}}(\phi\|\phi_{\mx})$ for $\alpha =0.1$ and $0.4$ when starting from the initial condition $\delta$, and for $\alpha =0.87$ when starting from the initial condition $\Gamma$. In the other three cases, however, the evolution of $\mathcal{D}_{\mathrm{KL}}(\phi\|\phi_{\mx})$ is monotonic. In contrast, the relative entropy $\mathcal{D}_{\mathrm{KL}}(\phi\|\phi_{\hcs})$ decays monotonically for the six cases, in qualitative agreement with the toy model.

\section{Velocity-inversion experiment}

A discussion about entropy is not complete if the issue of irreversibility is not included. In the case of \emph{elastic} hard disks, a simulated velocity-inversion experiment (produced by a sort of Maxwell's demon) was proposed more than forty years ago \cite{OB67,OB69,A71}, where schemes with ``anti-kinetic'' parts in the evolution were tested \cite{B67} and Loschmidt's paradox was discussed.
In Orban and Bellemans' pioneering works \cite{OB67,OB69}, during the evolution toward equilibrium the velocities of all elastic disks (simulated by MD) were inverted at a given waiting time $t_w$  and Boltzmann's $H$-functional was analyzed and seen to revert its decay by retracing its past values (anti-kinetic stage), in agreement with the underlying reversibility of the equations of motion. However, the $H$-functional resumed its decay after time $t=2t_w$ and, moreover, due to unavoidable error propagation \cite{KA04}, the initial value of $H$ was not exactly recovered if the velocity inversion took place after a sufficiently long waiting time. In a study involving irreversible particle dynamics, Aharony \cite{A71} observed that
the  anti-kinetic stage was not symmetric, the system rapidly forgetting the correlations
it had at $t_w$, and thereafter continuing to approach equilibrium.

In this section we revisit the velocity-inversion experiment in a freely cooling granular gas, modeled as inelastic hard spheres, where the collisional rules are given by \eqref{CR}. In this system, collisional symmetry is broken down by the inelasticity of collisions, closely related to a violation of microscopic reversibility. Consider  two colliding particles with precollision velocities $\{\vv_1,\vv_2\}$ and a relative orientation characterized by the unit vector $\s$ (with $\vv_{12}\cdot\s>0$). In that case, the postcollisional velocities are $\mathfrak{C}_{\s}\{\vv_1,\vv_2\}=\mathfrak{C}_{\s}\mathfrak{C}_{-\s}\{\vv_1^{\prime\prime},\vv_2^{\prime\prime}\}=\{\vv_1',\vv_2'\}$, where
\begin{equation}
    \vv_{1,2}^\prime = \vv_{1,2} \mp \frac{1+\alpha}{2}(\vv_{12}\cdot\s)\s, \quad \vv_{12}'\cdot\s=-{\alpha}\vv_{12}\cdot\s.
\end{equation}
Now, we invert the velocities $\{\vv_1',\vv_2'\}$ and obtain the subsequent postcollision velocities, $\mathfrak{C}_{\s}\{-\vv_1',-\vv_2'\}=\{-\vv_1^\dagger,-\vv_2^\dagger\}$, where
\begin{align}
\vv_{1,2}^\dagger=\vv_{1,2}\mp\frac{1-\alpha^2}{2}(\vv_{12}\cdot\s)\s,\quad \vv_{12}^\dagger\cdot\s=\alpha^2\vv_{12}\cdot\s.
\end{align}
Therefore, $\mathfrak{I}\mathfrak{C}_{\s}\mathfrak{I}\mathfrak{C}_{\s}\{\vv_1,\vv_2\}\neq \{\vv_1,\vv_2\}$ (where $\mathfrak{I}$ is the inversion-velocity operator) unless $\alpha=1$. We studied this effect from MD simulations in a computer experiment similar to those of the works discussed above \cite{OB67,OB69,A71}. A waiting time $s_w=0.5$ was chosen, several values of $\alpha$ were considered, and the evolution was monitored by $\mathcal{D}_{\mathrm{KL}}(\phi\|\phi_{\hcs})$, which plays the role of $H$ in the elastic case.

\begin{figure}[ht]
\centering
\includegraphics[width=8.cm,clip]{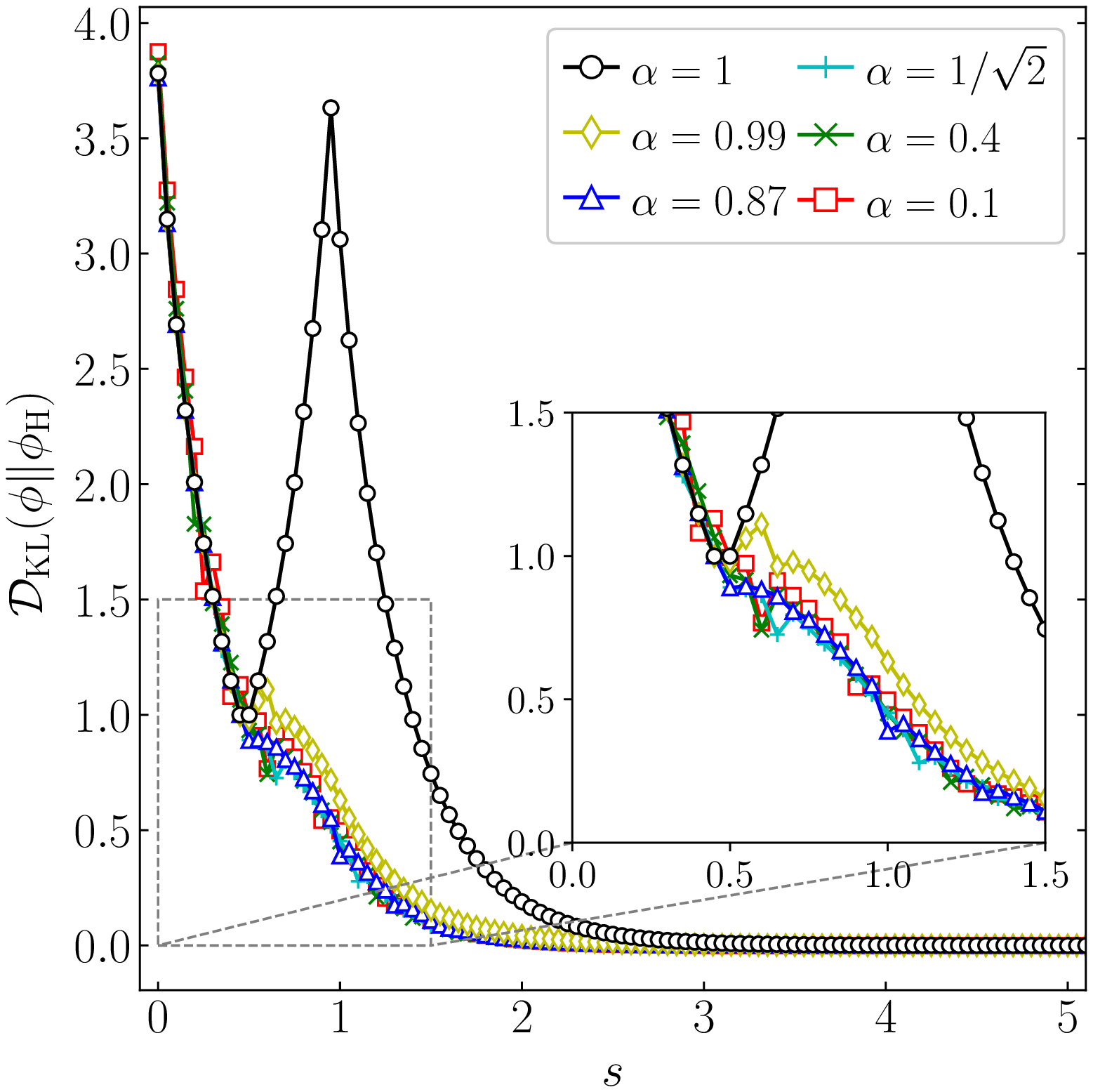}
\caption{Evolution of $\mathcal{D}_{\mathrm{KL}}(\phi\|\phi_{\hcs})$ in the velocity-inversion experiment (with a waiting time $s_w=0.5$). Symbols represent MD simulation results (joined with straight lines as a guide to the eye). The values of the coefficient of restitution are $\alpha=0.1$, $0.4$, $1/\sqrt{2}$, $0.87$, $0.99$, and $1$.
The inset magnifies the behavior around $s=0.5$. The error bars are smaller than the size of the symbols.}
\label{fig-2}       
\end{figure}

Figure \ref{fig-2} shows the time evolution of  $\mathcal{D}_{\text{KL}}(\phi\|\phi_{\hcs})$  when starting from the $\delta$ initial condition  and then applying the velocity inversion. The coefficients of restitution considered are $\alpha=0.1$, $0.4$, $1/\sqrt{2}$, $0.87$, $0.99$, and $1$. In the elastic case ($\alpha=1$), one recovers the results of Ref.\ \cite{OB67}, i.e., the system almost reaches the original configuration at $s=1$ but afterwards  it evolves toward equilibrium again. Whereas one expects that inelastic collisions erase completely the possibility of a reversible period, in the quasielastic case $\alpha=0.99$, although it is short, an anti-kinetic transient stage exists after the velocity inversion; this effect is translated into a small growth of $\mathcal{D}_{\text{KL}}(\phi\|\phi_{\hcs})$. Of course, the duration of the anti-kinetic regime becomes longer as $\alpha$ comes closer to $1$. On the other hand, as inelasticity increases ($\alpha\leq 0.87$), the influence of the velocity inversion is noticeable by a change of curvature only, and this  short effect shrinks with increasing inelasticity, as expected. The effect of inelasticity on the microscopic irreversiblity reflected by the behavior of $\mathcal{D}_{\text{KL}}(\phi\|\phi_{\hcs})$ is analogous to that observed by Aharony \cite{A71} for the conventional $H$-functional in the evolution toward equilibrium.

\section{Concluding remarks}

In this paper we have provided further evidence  from MD simulations on the conjecture that the Kullback--Leibler divergence $\mathcal{D}_{\mathrm{KL}}(\phi\|\phi_{\hcs})$ is a possible entropy-like functional for the case of isolated freely cooling granular gases \cite{GMMMRT15,MS20}, even for strongly inelastic systems.
Furthermore, this conjecture is supported by a simple toy model, which, on the other hand, predicts a nonmonotonic behavior of $\mathcal{D}_{\mathrm{KL}}(\phi\|\phi_{\mx})$ if $a_2(0)$ and $a_2^\hcs$ have opposite signs. This theoretical expectation has been nicely confirmed by our simulations.

Finally, the classical velocity-inversion experiment \cite{B67,OB67,OB69,A71}, originally devised for systems relaxing to equilibrium, has been applied on granular gases relaxing to the HCS and monitored via $\mathcal{D}_{\mathrm{KL}}(\phi\|\phi_\hcs)$. While, as expected, the initial configuration is almost perfectly recovered if the collisions are elastic ($\alpha=1$), microscopic reversibility is frustrated by inelasticity, no matter how small. In fact, a (short) anti-kinetic stage, where  $\partial_s \mathcal{D}_{\mathrm{KL}}(\phi\|\phi_\hcs)> 0$,   is only possible in the quasielastic regime (e.g., $\alpha=0.99$) and disappears for sufficiently high inelasticity ($\alpha\lesssim 0.9$).

\section*{Acknowledgements}
The authors acknowledge financial support from the Grant FIS2016-76359-P/AEI/10.13039/501100011033 and the Junta de Extremadura (Spain) through Grant No.\ GR18079, both partially financed by Fondo Europeo de Desarrollo Regional funds. A.M. is grateful to the Spanish Ministerio de Ciencia, Innovaci\'on y Universidades for a predoctoral fellowship FPU2018-3503.



%

%
%
%
%

\end{document}